\long\def\cut#1{}
\long\def\optional#1{}
\def\pagin#1{}
\def\boringfonts{y}   
\def\figflag{y} \input epsf  

%

\input harvmac  

\def\fonttest{y}

\ifx\boringfonts\fonttest
\else

\fi

\hyphenation{anom-aly anom-alies coun-ter-term coun-ter-terms
dif-feo-mor-phism dif-fer-en-tial super-dif-fer-en-tial dif-fer-en-tials
super-dif-fer-en-tials reparam-etrize param-etrize reparam-etriza-tion}


%
%
%
\newwrite\tocfile\global\newcount\tocno\global\tocno=1
\ifx\bigans\answ \def\tocline#1{\hbox to 320pt{\hbox to 45pt{}#1}}
\else\def\tocline#1{\line{#1}}\fi
\def\toclead{\leaders\hbox to 1em{\hss.\hss}\hfill}
\def\tnewsec#1#2{\newsec{#2}\xdef #1{\the\secno}%
\ifnum\tocno=1\immediate\openout\tocfile=toc.tmp\fi\global\advance\tocno
by1%
{\let\the=0\edef\next{\write\tocfile{\medskip\tocline{\secsym\ #2\toclead\the%
\count0}\smallskip}}\next}
\nobreak}
\def\tnewsubsec#1#2{\subsec{#2}\xdef #1{\the\secno.\the\subsecno}%
\ifnum\tocno=1\immediate\openout\tocfile=toc.tmp\fi\global\advance\tocno
by1%
{\let\the=0\edef\next{\write\tocfile{\tocline{ \ \secsym\the\subsecno\
#2\toclead\the\count0}}}\next}%
\nobreak}
\def\tappendix#1#2#3{\xdef #1{#2.}\appendix{#2}{#3}
\ifnum\tocno=1\immediate\openout\tocfile=toc.tmp\fi\global\advance\tocno
by1
{\let\the=0\edef\next{\write\tocfile{\tocline{ \ #2.
#3\toclead\the\count0}}}\next}
}
%
%

%
%
%
%

\gdef\prlmode{F}
%

%
\long\def\optional#1{}
\def\cmp#1{#1}         
%
%
\let\narrowequiv=\equiv
\def\equiv{\;\narrowequiv\;}

\fontdimen16\tensy=2.7pt\fontdimen17\tensy=2.7pt 



%
\def\ga{\gamma}

%
%

%
%
%
\def\boxit#1#2{
        $$\vcenter{\vbox{\hrule\hbox{\vrule\kern3pt\vbox{\kern3pt
	\hbox to #1truein{\hsize=#1truein\vbox{#2}}\kern3pt}\kern3pt\vrule}
        \hrule}}$$
}




%

\def\lfr#1#2{{\textstyle{#1\over#2}}} 



\def\splitexact#1#2{\mathrel{\mathop{\null{
\lower4pt\hbox{$\rightarrow$}\atop\raise4pt\hbox{$\leftarrow$}}}\limits
^{#1}_{#2}}}

%
%

%
%
%
%




%
%

\def\IM{isomorphism}

\def\eg{{\it e.g.}}\def\ie{{\it i.e.}}\def\etc{{\it etc.}}

%
%

\ifx\boringfonts\fonttest
\font\blackboard=cmssbx10 \font\blackboards=cmssbx10 at 7pt  
\font\blackboardss=cmssbx10 at 5pt
\else
\font\blackboard=msym10 \font\blackboards=msym7   
\font\blackboardss=msym5
\fi
\newfam\black
\textfont\black=\blackboard
\scriptfont\black=\blackboards
\scriptscriptfont\black=\blackboardss


%
\ifx\boringfonts\fonttest
\font\gothic=cmssbx10 \font\gothics=cmssbx10 at 7pt  
\font\gothicss=cmssbx10 at 5pt
\else
\font\gothic=eufm10 \font\gothics=eufm7
\font\gothicss=eufm5
\fi
\newfam\gothi
\textfont\gothi=\gothic
\scriptfont\gothi=\gothics
\scriptscriptfont\gothi=\gothicss

{\catcode`\@=11\gdef\oldcal{\fam\tw@}}
\newfam\curly
\ifx\boringfonts\fonttest\else
\font\curlyfont=eusm10 \font\curlyfonts=eusm7
\font\curlyfontss=eusm5
\textfont\curly=\curlyfont
\scriptfont\curly=\curlyfonts
\scriptscriptfont\curly=\curlyfontss

\fi
%

\ifx\boringfonts\fonttest\else\fi

\global\newcount\pnfigno \global\pnfigno=1
\newwrite\ffile
\def\pfig#1#2{Fig.~\the\pnfigno\pnfig#1{#2}}
\def\pnfig#1#2{\xdef#1{Fig. \the\pnfigno}%
\ifnum\pnfigno=1\immediate\openout\ffile=figs.tmp\fi%
\immediate\write\ffile{\noexpand\item{\noexpand#1\ }#2}%
\global\advance\pnfigno by1}

%
%
\def\tfig#1{Fig.~\the\pnfigno\xdef#1{Fig.~\the\pnfigno}\global\advance\pnfigno
by1}

%
%
%
%
\def\figI{y}
\def\ifigure#1#2#3#4{
\midinsert
\ifx\figflag\figI
 \ifx\htflag\figI
 \vbox{
  \href{file:#3}
{Click here for enlarged figure.}}
 \fi
 \vbox to #4truein{
 \vfil\centerline{\epsfysize=#4truein\epsfbox{#3}}}
\else
\vbox to .2truein{}
\fi
\narrower\narrower\noindent{\bf #1:} #2
\endinsert
}








%
%

%


\def\inbar{\,\vrule height1.5ex width.4pt depth0pt}
\def\IB{\relax{\rm I\kern-.18em B}}
\def\IC{\relax\hbox{$\inbar\kern-.3em{\rm C}$}}
\def\ID{\relax{\rm I\kern-.18em D}}
\def\IE{\relax{\rm I\kern-.18em E}}
\def\IF{\relax{\rm I\kern-.18em F}}
\def\IG{\relax\hbox{$\inbar\kern-.3em{\rm G}$}}
\def\IH{\relax{\rm I\kern-.18em H}}
\def\II{\relax{\rm I\kern-.18em I}}
\def\IK{\relax{\rm I\kern-.18em K}}
\def\IL{\relax{\rm I\kern-.18em L}}
\def\IM{\relax{\rm I\kern-.18em M}}
\def\IN{\relax{\rm I\kern-.18em N}}
\def\IO{\relax\hbox{$\inbar\kern-.3em{\rm O}$}}
\def\IP{\relax{\rm I\kern-.18em P}}
\def\IQ{\relax\hbox{$\inbar\kern-.3em{\rm Q}$}}
\def\IR{\relax{\rm I\kern-.18em R}}
\font\cmss=cmss10 \font\cmsss=cmss10 at 10truept
\def\IZ{\relax\ifmmode\mathchoice
{\hbox{\cmss Z\kern-.4em Z}}{\hbox{\cmss Z\kern-.4em Z}}
{\lower.9pt\hbox{\cmsss Z\kern-.36em Z}}
{\lower1.2pt\hbox{\cmsss Z\kern-.36em Z}}\else{\cmss Z\kern-.4em Z}\fi}
\def\IGa{\relax\hbox{${\rm I}\kern-.18em\Gamma$}}
\def\IPi{\relax\hbox{${\rm I}\kern-.18em\Pi$}}
\def\ITh{\relax\hbox{$\inbar\kern-.3em\Theta$}}
\def\IOm{\relax\hbox{$\inbar\kern-3.00pt\Omega$}}

\def\micron{$\mu$m}

\def\ee#1{\cdot 10^{#1}}
\def\eem#1{\cdot 10^{-#1}}


\long\def\suppress#1{}
\suppress{\def\boringfonts{y}  

\baselineskip=20pt
\def\ifigure#1#2#3#4{\nfig\dumfig{#2}}

} 


\def\testp{T}

\Title{\vbox{\hbox{UPR--704T}}}{\vbox{\centerline{Spontaneous Expulsion of
Giant
Lipid }
\vskip2pt\centerline{Vesicles Induced by Laser Tweezers}
}}

\centerline{J. David Moroz and Philip Nelson}\smallskip
\centerline{Department of Physics and Astronomy, University of Pennsylvania}
\centerline{Philadelphia, PA 19104 USA}
\medskip
\centerline{Roy Bar-Ziv and Elisha Moses}\smallskip
\centerline{Department of Physics of Complex Systems, }
\centerline{Weizmann Institute of Science, Rehovot 76100,
Israel}
\bigskip

\ifx\answ\bigans \else\noblackbox\fi

Irradiation of a giant unilamellar lipid bilayer vesicle with a focused
laser spot leads to
a tense pressurized state which persists indefinitely after laser
shutoff.  If the vesicle
contains another object it can then be
gently and continuously expelled from the tense outer vesicle.
Remarkably, the inner object can be almost as large as
the parent vesicle; its volume is replaced during
the exit process. We offer a qualitative theoretical model to explain
these and related phenomena. The main hypothesis is that the laser
trap pulls in lipid and ejects it in the form of submicron objects,
whose osmotic activity then drives the expulsion.

\Date{7/96
}\noblackbox


\def\micron{$\,\mu$m}
\def\mus{$\,\mu$m$^2$}
\def\muc{$\,\mu$m$^3$}

  \def\laser{_{\rm laser}}

\let\thi=\thinspace
\def\ee#1{\cdot 10^{#1}}
\def\eem#1{\cdot 10^{-#1}}
\def\UT{{\rm \,dyn/cm}}
\def\UP{{\rm \,dyn/cm^2}}

\hfuzz=3truept


\lref\rKaCu{A.                   Katchalsky and P. Curran, {\sl
Nonequilibrium thermodynamics in biophysics} (Harvard University
Press, 1965).}
\lref\rLLfluids{L. Landau and E. Lifshitz, {\sl Fluid mechanics}
(Pergamon, 1959). Actually the classical exact solution for
diverging and converging
flow between rigid walls underestimates $Q$, since the walls
are really fluctuating fluid membranes.}
\lref\rLiLiSa{R. Lipowsky, \cmp{``Generic interactions of flexible
membranes,''} in {\sl Handbook of
biological physics,} ed. R. Lipowsky and E. Sackmann (Elsevier, 1995)
pp.~548, 554.}%
\lref\rudorev{U. Seifert, ``Configurations of fluid membranes and
vesicles," Advances in Physics, in press}
\lref\rHelsteric{W. Helfrich, Z. Naturforsch. {\bf33a} (1978) 305.}
\lref\rOsPe{The microscopic origin of
osmotic flow has recently been explained in G. Oster and C. Peskin,
\cmp{``Dynamics of osmotic fluid flow,''} in {\sl  Mechanics of swelling:
from clays to living cells and tissues} ed. T. K. Karalis
(Springer, 1992) pp. 731-742.}
\lref\rNeZh{D. Needham and D. Zhelev, ``Tension-stabilized pores in
giant vesicles: determination of pore size and pore line tension,''
Biochim. et Biophys. Acta {\bf1147} (1993) 89\pagin{--104}.}
\lref\rTaupin{C. Taupin, M. Dvolaitzky, and C. Sauterey, \cmp{``Osmotic
pressure induced pores in phospholipid vesicles,''} Biochemistry
{\bf14} (1975) 4771\pagin{--4775}.}
\lref\rpassages{D. Roux; G. Porte ??}
\lref\rMeGa{\cut{M. Dvloaitzky, M. Guedeau-Boudeville, and L. L\'eger,
\cmp{``Polymerization-induced shrinkage in giant butadienic lipid
vesicles,''} Langmuir {\bf8} (1992) 2595\pagin{--2597};}
F. Menger and K. Gabrielson, \cmp{``Chemically-induced birthing
and foraging in vesicle systems,''} J. Am. Chem. Soc. {\bf116} (1994)
1567\pagin{--1568}.}
\def\rEvAdh{}
\lref\rHelfAdh{E. Evans, \cmp{``Detailed mechanics of membrane-membrane
adhesion and separation,''} Biophys. J. {\bf 48} (1985)
175\pagin{--183}; W. Helfrich, \cmp{``Tension-induced mutual adhesion and a
conjectured superstructure of lipid membranes,''} in {\sl Handbook of
biological physics,} ed. R. Lipowsky and E. Sackmann (Elsevier, 1995)
pp.~691--721.}%
\lref\EvYe{E. Evans and A. Yeung, \cmp{``Hidden dynamics in rapid
changes of bilayer shape,''} Chem. Phys. Lipids, 73 (1994) 39--56.}%
\lref\rHeSeUSeff{W. Helfrich and R. Servuss, ``Undulations, steric
interaction and cohesion of fluid membranes,'' Nuovo Cimento {\bf3D}
(1984) 137; U. Seifert, ``The concept of effective tension for
fluctuating vesicles,'' Z. Phys. {\bf B97} (1995) 299\pagin{--309}.}
\lref\rEvRa{E. Evans and W. Rawicz, \cmp{``Entropy-driven tension and
bending elasticity in condensed-fluid membranes,''} Phys. Rev. Lett.
{\bf64} (1990) 2094.}
\lref\rGoSt{J. Goll and G. Stock, ``Determination by photon
correlation spectroscopy of particle size distributions in lipid
vesicle suspensions,'' Biophys. J. {\bf19} (1977) 265\pagin{--273};
J. Goll, F. Carlson, Y. Barenholz, B. Litman, and T. Thompson,
``Photon correlation spectroscopic study of the size distribution of
phospholipid vesicles,'' Biophys. J. {\bf38} (1982) 7\pagin{--13}.}

\lref\rChern{\optional{L. Chernomordik, M. Kozlov, G. Melikyan, I. Abidor,
V. Markin,  and Y. Chizmadzhev, ``The shape of lipid molecules and
monolayer membrane fusion,'' Biochim. Bioph. Acta {\bf812} (1985)
643\pagin{--655}; S. Leikin, M. Kozlov, L. Chernomordik,
V. Markin,  and Y. Chizmadzhev, \cmp{``Membrane fusion,''}
J. Theor. Biol. {\bf129} (1987) 411\pagin{--425};
}L. Chernomordik, G. Melikyan, and Y. Chizmadzhev,
\cmp{``Biomembrane fusion,''} Biochim. Bioph. Acta {\bf906} (1987)
309\pagin{--352}.}
\lref\rRaPa{R. Rand and V. Parsegian, ``Mimicry and mechanism in
phospholipid models of membrane fusion,'' Ann. Rev. Physiol. {\bf48}
(1986) 201\pagin{--212}.}
\lref\rHaHe{W. Harbich and W. Helfrich, \cmp{``Alignment and opening of giant
lecithin vesicles by electric fields,''} Z. Naturforsch. {\bf34a}
(1979) 1062\pagin{--1065}.}
\lref\rSowersa{L. Boni and S. Hui, \cmp{``The mechanism of PEG-induced
fusion in model membranes,''} in {\sl Cell Fusion}
ed. A. Sowers (Plenum, 1987) pp.~301--330.}
\lref\rSowersb{E. Schierenberg, \cmp{``Laser-induced cell fusion,''}
in {\sl Cell Fusion} ed. A. Sowers (Plenum, 1987) pp.~409--418.}
\lref\rHap{J. Happel and H. Brenner, {\sl Low Reynolds-number
hydrodynamics} (Prentice-Hall, 1965).}
\lref\rZhNeb{D. Zhelev and D. Needham, \cmp{``The influence of
electric fields on biological and model membranes,''} in {\sl
Biological effects of
electric and magnetic fields} ed. D. Carpenter and S. Ayrapetyan
(Academic, 1994) pp.~105--142.}
\lref\rBMMS{R. Bar-Ziv, R. Menes, E. Moses, and S. Safran, \cmp{``Local
unbinding of pinched membranes,''}
Phys. Rev. Lett. {\bf 75} (1995) 3356.}
\lref\rpearla{R. Bar-Ziv and E. Moses, \cmp{``Instability and `pearling'
states produced in tubular membranes by competition of curvature and
tension,''} Phys. Rev. Lett., {\bf73} (1994) 1392.}%
\lref\rBFM{R. Bar-Ziv, T. Frisch, and E. Moses, \cmp{``Entropic expulsion in
vesicles,''} Phys. Rev. Lett. {\bf 75} (1995) 3481\pagin{--3484}.}
\lref\rGNPS{P. Nelson, T. Powers, and U. Seifert, \cmp{``Dynamic theory of
pearling instability in cylindrical vesicles,''}
Phys. Rev. Lett. {\bf74} (1995) 3384\pagin{-3387};
R. Goldstein, P. Nelson, T. Powers, and U. Seifert,
\cmp{``Front propagation and the pearling instability of tubular
vesicles,''} J. Physique II
(France) {\bf6} (1996) 767\pagin{--796}.}
\lref\rGrOl{R. Granek and Z. Olami, \cmp{``Dynamics of Rayleigh-like
instability induced by laser tweezers in tubular vesicles of
self-assembled membranes,''} J. Phys. II France {\bf5} (1995)
1348\pagin{--1370}.}
\def\rpearlth{\rGNPS\rGrOl}
\lref\rFink{A. Finkelstein, {\sl Water movement through lipid
bilayers, pores, and plasma membranes\cut{: theory and reality},} (Wiley,
1987). }
\lref\rHeSe{W. Helfrich and R. Servuss, ``Undulations, steric
interaction and cohesion of fluid membranes,'' Nuovo Cimento {\bf3D}
(1984) 137.}
\lref\rAlberts{B. Alberts, D. Bray, J. Lewis, M. Raff, K. Roberts, and
J. Watson, {\sl Molecular biology of the cell}
(Garland, 1989).}
\lref\rmica{R. Pashley and J. Israelachvili, ``Hydration forces
between mica surfaces in aqueous electrolyte solutions,''
J. Colloid. Interface Sci. {\bf80} (1981) 153\pagin{--162}.}
\lref\rNeePerm{D. Needham, in {\sl Permeability and stability of lipid
bilayers,} ed. E. Disalvo and
   S. Simon (CRC Press, 1995) p.~69.}
\lref\rMarsh{D.Marsh, {\sl CRC handbook of lipid bilayers} (CRC Press,
1990).}
\lref\rKaCu{A.                   Katchalsky and P. Curran, {\sl
Nonequilibrium thermodynamics in biophysics} (Harvard University
Press, 1965).}
\lref\rBean{C. Bean, ``The physics of porous membranes,'' in {\sl
Membranes --- macroscopic systems and models} ed. G. Eisenman (Marcel
Dekker, 1972).}
\lref\rRSS{W. Russell, D. Saville, and W. Schowalter, {\sl Colloidal
dispersions} (Cambridge, 1989)\optional{ p. 50}.}
\lref\rfuse{E. Chang, ``Pressure as a probe of vesicle fusion,'' in {\sl Cell
 Fusion}
ed. A. Sowers (Plenum, 1987) pp.~353--364.}

\newsec{Introduction}
Nature has carefully designed lipid molecules to yield extraordinary
material properties. In water, lipids spontaneously self-assemble to
form bilayer membranes, loose fluid confederations of molecules which
nevertheless offer tremendous resistance to mechanical disruption,
topology change, and permeation \rAlberts. For example, despite a
thickness of only a few nanometers pure bilayer membranes withstand surface
tensions of more than 1~dyn/cm before rupture \rEvRa.
Of course the inert character of membranes is crucial for their role
in cells as tough, flexible partitions. Cells have specific machinery
to induce fusion, pore formation, \etc\ only when required.
Understanding this machinery, and finding new artificial mechanisms
for controlled bilayer reorganization, are key tasks for cell
biology. {In this paper we describe a new technique for selectively
disrupting membranes using laser tweezers.}

The development of laser tweezer technology has opened the door to the
direct manipulation of micron-scale objects with adjustable piconewton
scale forces. Two of us have applied tweezers directly to bilayer
membranes to generate a number of striking phenomena, including
dynamic shape transformations and membrane unbinding \rpearla\rBMMS.
A theoretical understanding has begun to emerge in which a primary
effect of the laser is to create sudden {\it surface tension} $\Sigma$
in the bilayer, in the regime $10^{-4}$--$10^{-3}\UT$ and controlled
by the applied laser power \rpearlth.

Perhaps most striking of all,
laser tweezers can set up conditions which persist long after the
laser is shut off and which lead 
to the {\it spontaneous expulsion} of interior objects from a vesicle
without otherwise damaging either the ``parent'' or the ``daughter''
object \rBFM\ (\tfig\fone). In the light of our earlier remarks it is
remarkable
that this dramatic membrane reorganization should happen for tensions
a thousand times smaller than what is normally
required. We will describe the observed phenomena and sketch {a
proposed mechanism}.

\ifigure\fone{Typical spontaneous
giant expulsion event. Selected video
frames are shown, separated by 0.12\thi s.
A large DMPC vesicle of radius $R=4.5$\micron\ in pure water (see
text) initially contains a
smaller daughter of radius $r=3.3$\micron. The temperature was
constant at $31^\circ$C. The laser was focused to a spot about
0.3\micron\ in diameter; an intensity of 6\thi mW was measured through
this spot.
After 22\thi s of tweezing the laser was shut off. Then
the inner vesicle adhered to the outer one, waited 12\thi s, and
finally emerged as shown. In the final frames the daughter vesicle
emerges rapidly, leaving the focal plane. The dark object trapped in
the daughter vesicle plays no part in the expulsion process.
Scale bar is $10 \mu$m in the horizontal direction; in the vertical
direction the same bar is $10.6 \mu$m, due to an asymmetry in our
camera. }{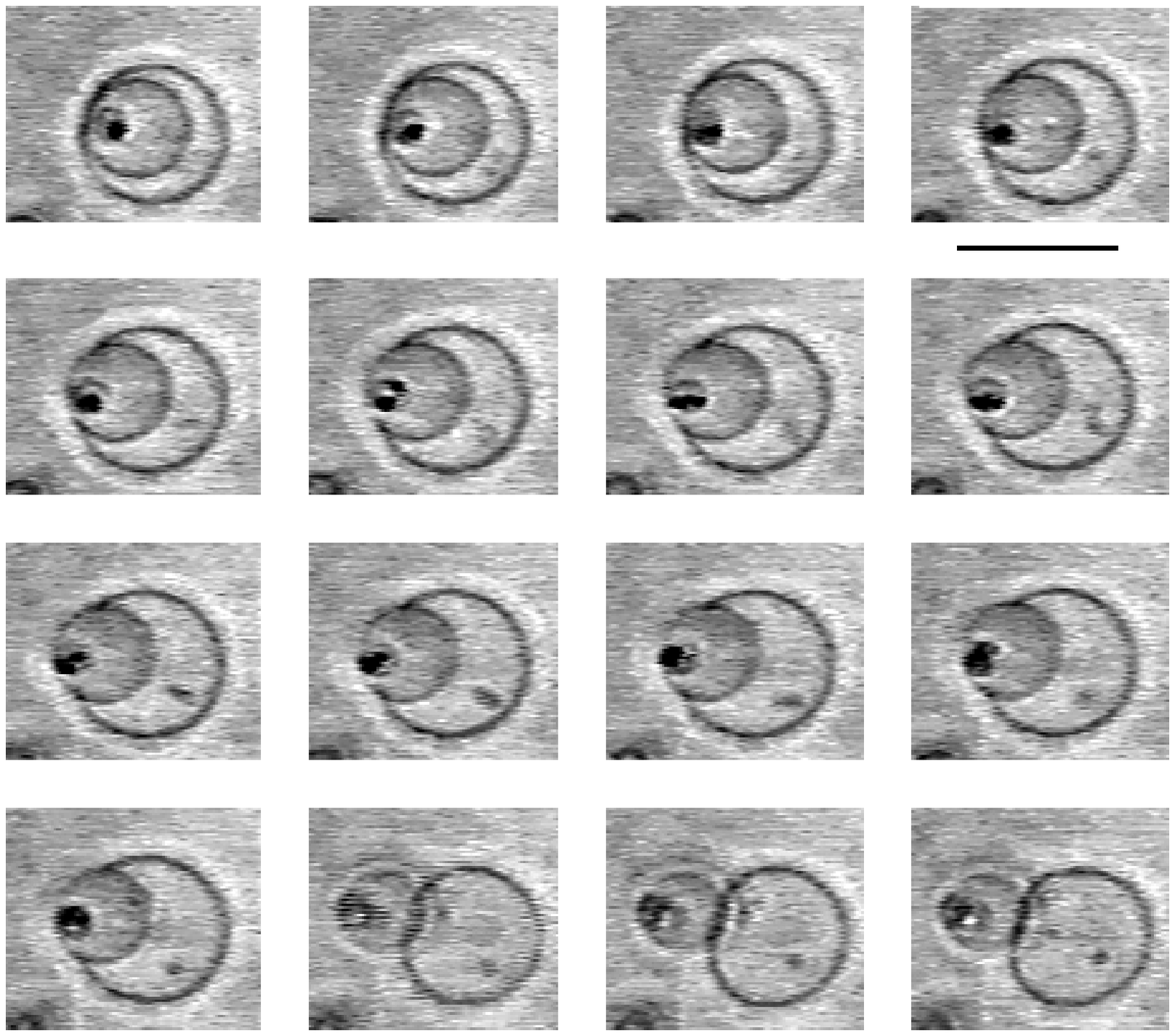}{5}

Other authors have reported that vesicle expulsion can also
be induced by chemical means \rMeGa. The phenomena we will describe
are triggered solely by laser action.
Related phenomena are also seen in vesicles tensed using the
micropipette method or osmotic shock \ref\rpriv{D. Fygenson, private
communication.}, but these
events  require far greater tensions than the ones
reported here.
Finally, our work extends initial observations described in
\rBFM\ in several ways which proved crucial to obtaining the
theoretical picture introduced here.

\newsec{Methods}
We report results based on vesicles produced from
dimyristoyl--phosphatidylcholine (DMPC, purchased from Sigma) using
standard protocols
\ref\protocol{D. Needham and  E. Evans, Biochemistry, {\bf 27} (1988) 8261;  H.
 P.
Duwe, J. Kas and E. Sackmann, J. Phys. (France) {\bf 51} (1990) 945;  M.
Mutz and W.
Helfrich, J. Phys. (France) {\bf 51} (1990) 991.}.
We have obtained similar results with stearoyl-oleoyl--phosphatidylcholine
(SOPC)
and digalactosyl--diglyeceride (DGDG) (Sigma), all of them uncharged,
zwitterionic
lipids. The experiments were performed in the fluid state of the membranes
in a closed
cell. In DMPC we varied the
temperature
from $26^\circ$C (just above the fluid transition) up to
$85^\circ$C  and observed expulsion in the full temperature range.
We constrained the
vesicles' initial volume osmotically.
We used glucose concentrations ranging from 0M (we estimate that up to
1\thi mM of uncontrolled impurities are always present in the
solution) to 0.5\thi M, and
consistently obtained expulsion. Even 1\thi mM is sufficient to
``clamp'' the volume to high accuracy.

Our experimental apparatus uses an inverted optical microscope (Zeiss
Axiovert 135TV
with Planapochromat Ph3, 63x, NA 1.4 objective), in a commonly used optical
tweezers
setup
\ref\rBlockSim{S.M. Block, in {\sl Noninvasive techniques in
cell biology},
 375-402, (Wiley-Liss, 1990); K. Svoboda and S. M. Block, Annu.
Rev. Biophys. Biomol. Struct., {\bf 23} (1994) 247; A. Simon, Ph.D
Thesis, Univ. of Chicago, (June, 1992).}. A significant point lies in using
the 488-514 nm Ar ion laser beam (Coherent, Innova 70) to produce the
optical trap.
This creates a tighter beam waist than infrared lasers and therefore a stronger
electromagnetic field. The laser power ranged
from 10 to 100 mW. Heating at these power inputs is estimated to be lower than
$\Delta T\approx 0.5$K~{\rBlockSim\rBMMS}.

Tension is produced in the membrane because the lipid has a higher
refractive index
than water and is pulled by the high electromagnetic field  into
the optical
trap~\rGNPS. {The  electrostatic energy difference per unit area
in the laser spot gives an estimate for the surface tension transmitted to the
rest of the membrane, on the order of $\Sigma_{\rm laser}
\approx10^{-3}\,\UT$ for a power
input of 25 mW.  For comparison, a flaccid vesicle of
radius $R=10$\micron\ has $\Sigma$ in the range of
$\kappa/R^2\approx10^{-6}\,\UT$,} where $\kappa$ is the bending stiffness.

As the laser is applied to a flaccid vesicle, it gradually loses its
fluctuations and
becomes round and taut, a directly observable sign of tension in the
membrane~{\rBFM}. We have measured
surface area losses on the order of $10\%$ prior to expulsion, though
other experiments show that much larger amounts can
disappear.

\newsec{Observations}
\fone\ shows a typical giant, spontaneous expulsion
event.\foot{Small spontaneous expulsions, $r/R\le 0.1$, were
reported in \rBFM. The inflow described below was not visible in this
regime.  Also {\it stimulated} expulsions, where the laser was never
turned off, have been seen for 
double-bilayer vesicles~\rBFM. Here we study only
spontaneous, giant expulsion.} For concreteness we will focus our
analysis on
this event, but the same qualitative physics appears in many similar
events. \tfig\ftwo\ defines notation. First the
inner vesicle was drawn tense by tweezing for several seconds.
The gradual increase in surface tension was clearly visible from
the gradual suppression of visible thermal motion. {In other
experiments we
have found that }vesicles pressurized in this way remain tense
indefinitely (at least hours), and that during tweezing
the vesicle volume remained approximately fixed or grew slightly,
while its area
decreased gradually by about 10\%. Next the same procedure was repeated
for the outer vesicle. Then the laser was shut off.

Often the tense inner vesicle wandered for some seconds in Brownian
motion before encountering the outer wall. Upon close enough approach,
adhesive forces snapped the vesicles together rapidly, sometimes
visibly deforming the outer vesicle towards the inner before the two
could draw together.  Strong adhesive forces comparable to the applied
membrane tension are expected on experimental and theoretical grounds
\rEvAdh\rHelfAdh.

\ifigure\ftwo{Notation used in the text.}{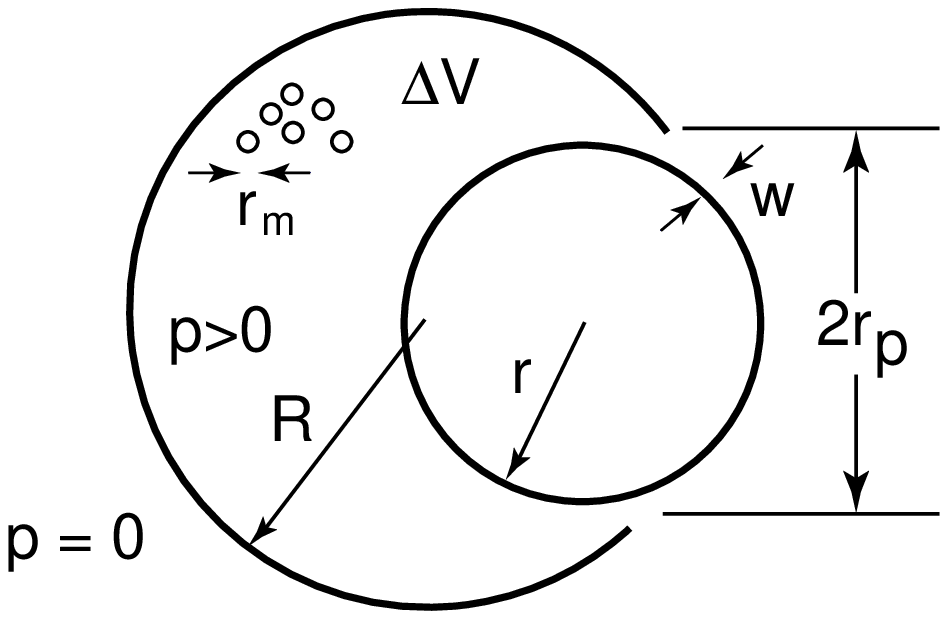}{3}

After a waiting period, {typically 10 seconds, }the inner
vesicle began to emerge gradually.
The point of initial tweezing was generally
not related to the point of the subsequent exit.
Though the daughter vesicle's volume was sometimes as
large as 40\% that of the parent, the parent always remained tense
(\ie\ spherical and nonfluctuating) through the halfway point. Most
events slowed considerably near the halfway point, and some stalled
and retracted prior to this. Every event which passed the halfway
point completed rapidly, in one or two video frames.

In the final state the  daughter was fully detached and could be readily
pulled away by the usual tweezer manipulation. Throughout the process
the volume and area of the daughter remained roughly constant
(\tfig\fthree a), from which we infer that at least one of its
monolayer walls was intact throughout.
Remarkably, the area of the {\it outer} vesicle also remained constant
through the halfway point (\fthree a). Thus not only was the
final surface area equal to the original, but at every intermediate
step the outer vesicle area (excluding the absent cap where the
daughter is emerging) remained constant. Correspondingly the volume
$\Delta V$ of
the space between the vesicles was {\it not} constant but rather grew
(\fthree b).

\ifx\figflag\figI
\midinsert
 \vbox to 2.8truein{
 \vfil\centerline{\hbox{\epsfysize=2.8truein\epsfbox{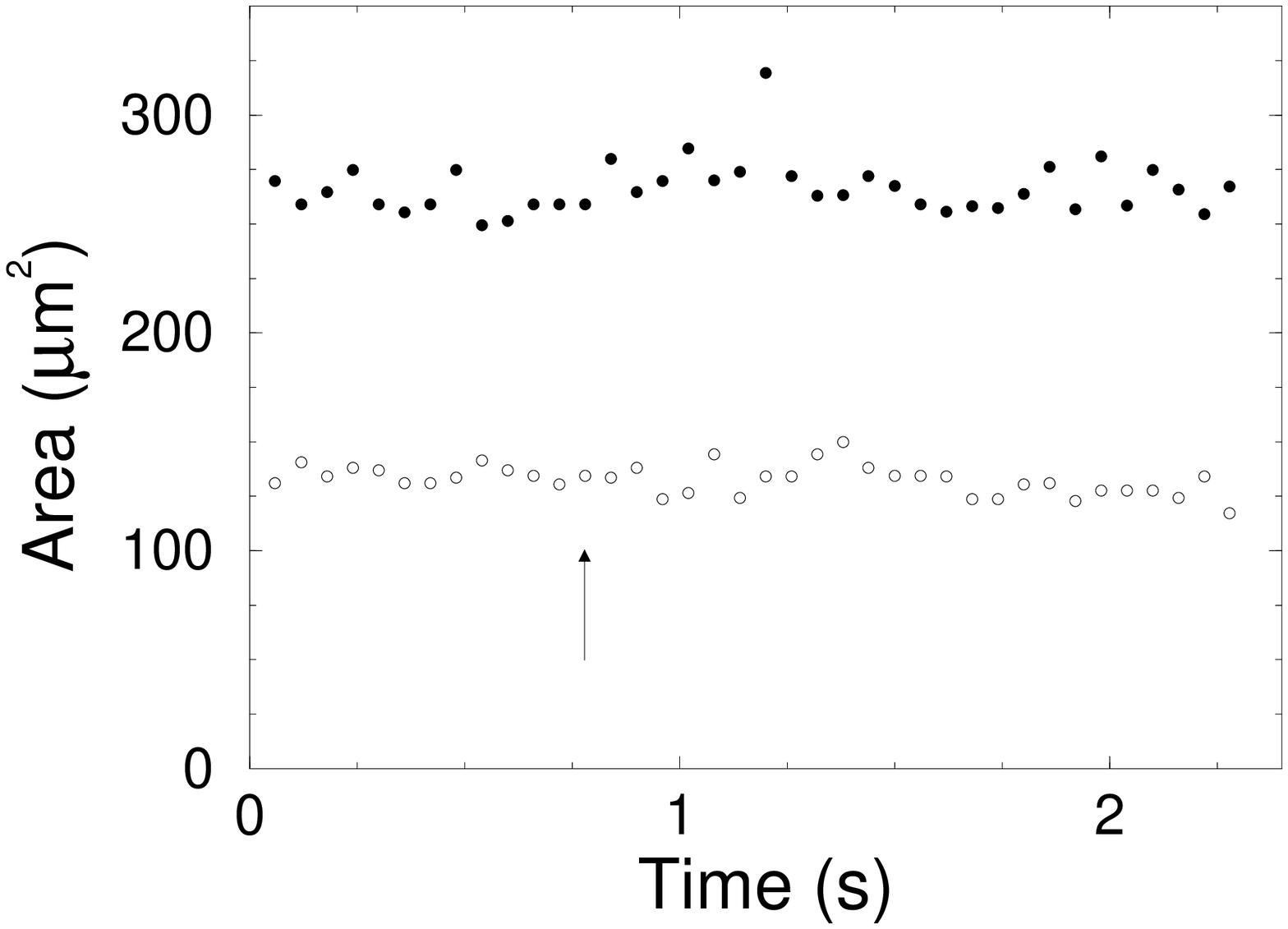}}
\hskip-.1truein\hbox{\epsfysize=2.8truein\epsfbox{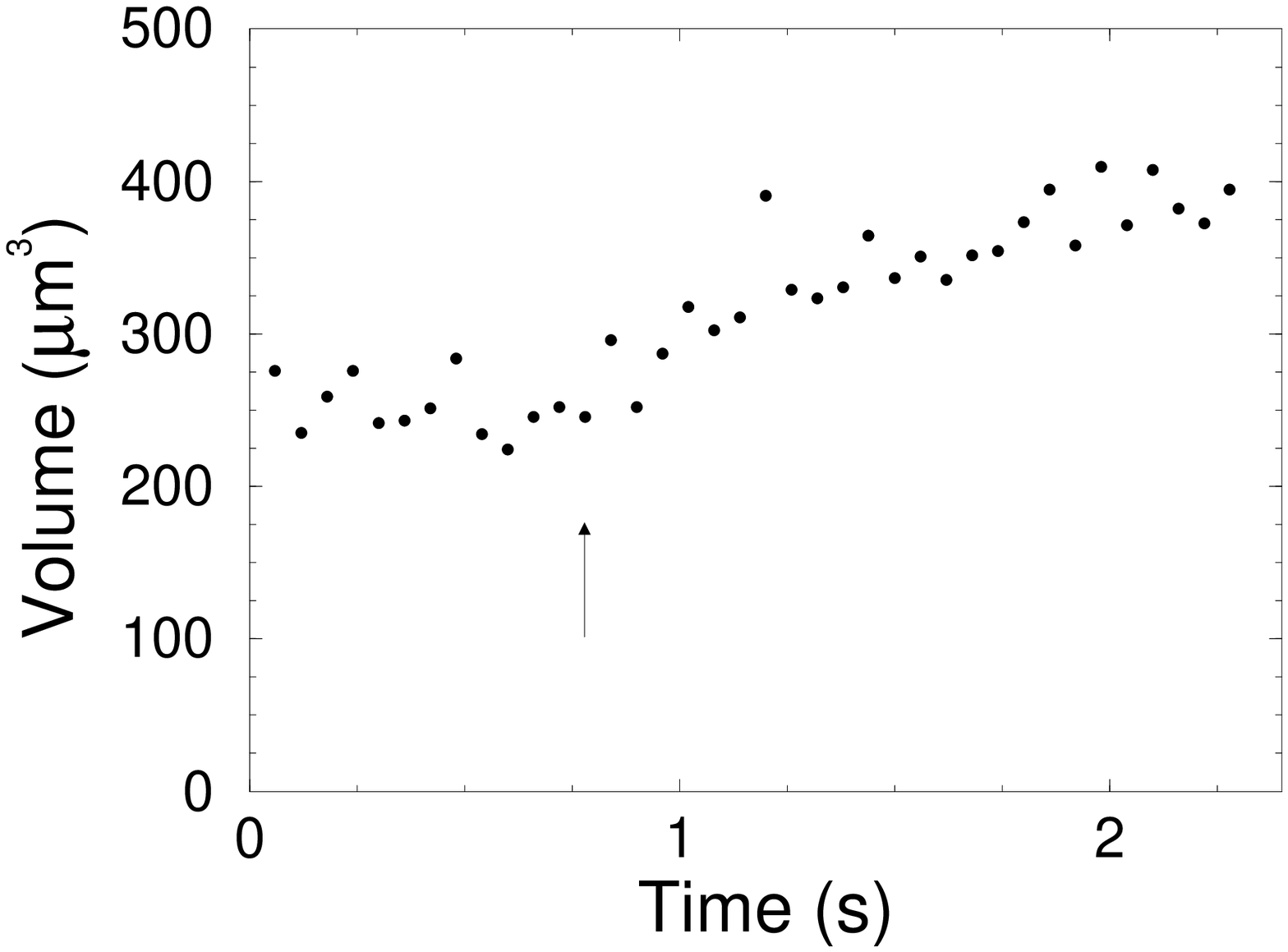}}}}
\smallskip\narrower\narrower\noindent{\bf \fthree:} (a)~Surface area of
inner and outer vesicles as a function of time. (b)~Volume between
inner and outer vesicles. The arrow refers to the first frame shown in
\fone; the points are spaced at intervals of 0.06\thi s.
\endinsert
\else
\ifigure\fthree{(a)~Surface area of
inner (open dots) and outer (solid dots) vesicles as a function of
time. (b)~Volume between
inner and outer vesicles. The arrow refers to the first frame shown in
\fone; the points are spaced at intervals of 0.06\thi s.}{}{}
\fi

\newsec{Model}
One reason why spontaneous giant expulsion is so surprising is that
after the laser is shut off, considerable
energy must be stored somewhere in order to push open the exit pore, and
yet there is no obvious elastic element in the system. The water is
essentially incompressible,
and the membrane itself can only stretch by a
couple of percent, even when we account for entropic elasticity
\rEvRa. Where is the ``spring'' driving expulsion?

A second mystery is that whatever the thermodynamic force driving
expulsion, it must overcome a nucleation energy barrier arising from
the line energy $\ga_0$ for creating  pores. DMPC membranes
withstand surface tensions of a few $\UT$ before rupture, implying that
$\ga_0$ is a few times $10^{-7}\,$dyn (see \rTaupin), and indeed
vesicle expulsion without laser action (for
example by micropipette aspiration \rpriv) requires tensions of this
magnitude. The laser-induced tension is much less than this, and
indeed tweezed vesicles without large included objects do remain
tense indefinitely. What laser-induced
mechanism creates an exit pore at low tension, but only in the presence of
an interior object?

We get  clues to both these puzzles when we note that
the direct observation of membrane tension $\Sigma$ implies a
corresponding hydrostatic pressure $p=2\Sigma/R$ inside the large
vesicle, and yet water is clearly seen to be {\it entering,} not leaving,
the intermembrane space. This implies an {\it osmotic flow} mechanism:
an excess of some solute in the interior maintains $p$ while pulling
more water in to dilute the solute.
Such a mechanism may at first sound
paradoxical: with up to an atmosphere of osmotic pressure due to
sugar on both sides of the outer vesicle, significant volume change
would seem to require an absurdly large initial pressure difference.
Indeed nothing of the sort can happen if pure water enters by
permeation through the bilayer. (Moreover the known permeation rate of
{70\micron/sec for DMPC} \rNeePerm\ is much too small to give the
observed influx.)
If however a gap of width $w$ much larger than a sugar
molecule opens (see \ftwo), then sugar becomes osmotically
irrelevant. Indeed as mentioned a wide range of sugar concentrations
has little effect on expulsion. We will return later to the origin of
the gap.

Thus we must identify some other osmotically-active solute driving
expulsion. For this we focus on the area loss during the initial laser
tweezing. Roughly 10\% of the original 250\mus\ of surface
disappears permanently, leaving behind no visible scar or buildup. Our
main physical hypothesis is that membrane
area lost during the initial tweezing gets packaged into a suspension
of small, optically unresolvable objects of size $r_m$, some of which
get trapped inside the outer vesicle. This suspension is
the ``spring'' we were looking for at
the start of this section: its osmotic pressure is what stores the
energy needed to open the exit
pore.

These objects could be very small vesicles or membrane fragments.
Perhaps more likely, the intense electric field in the laser spot
may break apart the lipid molecules, creating a new chemical species
analogous to the surfactant introduced directly by Menger and Gabrielson
\rMeGa.\foot{We thank J.\ Israelachvili, U.\ Seifert, and D.\
Zhelev for independently suggesting this mechanism to us.} Lipids are well
known to be fragile; they degrade into lysolipids with heat
or even the passage of time, with a significant
reduction in their line energy \ref\rHaHe{W. Harbich and W. Helfrich,
\cmp{``Alignment and opening of giant
lecithin vesicles by electric fields,''} Z. Naturforsch. {\bf34a}
(1979) 1062\pagin{--1065}.}. The new species could then
stabilize very small micelles, which would then avoid rejoining the
membrane due to hydrodynamic interactions. Thus our hypothesis also
explains the
permanent area loss upon tweezing. Further work will be required
to measure the size $r_m$. For concreteness we will illustrate our
mechanism using the
smallest reasonable value, $r_m=5\,$nm.
Once formed, the micelles can easily escape the laser
trap. Their trapping energy is proportional to their area,
$E\approx\Sigma\laser\cdot4\pi{r_m}^2$, but their thermal energy
$k_BT$ is much larger than this, so they readily diffuse away, making
room for more material to enter the trap.

Let us estimate the osmotic pressure $\Delta\pi$ of the micelles. The
lost membrane occupied volume 25\mus$\cdot4\,$nm. Subdividing into $N$
micelles of volume $\lfr{4\pi}3 {r_m}^3$ gives $N\approx
2\ee5$. Supposing that half of these remain trapped in volume $\Delta
V\approx250$\muc\ while the other half escape, the
volume fraction is then $4\eem4$. We may thus use {the  ideal gas
formula (van 't~Hoff's law) }to
get $\Delta\pi\approx k_BTN/2\Delta V=17\UP$.

In the presence of our hypothetical small micelles,
the line tension is no longer a constant. The rapid jump to adhesion
can trap micelles between the two vesicle walls. Eventually a micelle
can incorporate into the outer wall, delivering its modified lipid
contents and greatly decreasing the energy required to form an edge
\rChern. Thus we overcome the nucleation energy barrier, but only when
an inner object sticks to the outer vesicle wall, as observed. As the
perimeter of the pore increases, and as the
impurities diffuse away from the edge, the effective $\ga$ rises towards
its nominal value $\ga_0$. For a direct estimate of $\ga_0$ in our
experiment, we can examine the rapid completion stage of
expulsion. Here the exit
pore snaps shut, propelling the inner vesicle a few microns in one or
two video frames. From \fone\ we estimate a speed $v\approx
3\eem3$cm/s. Setting this equal to the Stokes drag on a sphere of
radius $r=3.3$\micron\ gives the order-of-magnitude estimate for the
line energy $\ga_0\approx10^{-7}\,$dyn. Other events gave similar values.

Once the exit pore is established, we have argued that it should not
seal tightly around the daughter vesicle but rather must leave a gap
of width $w$. Indeed while the line energy $\ga_0$ tries to close the
gap, thermal fluctuations constantly keep it open. We can
estimate the average width $w$ of the gap by adapting  Helfrich's
``steric repulsion'' argument \rHelsteric\ (see \rLiLiSa). At the
halfway point of
expulsion, the rim of the exit pore is a tense fluctuating line, and
so it feels an effective repulsive free energy of $1.89\cdot{2\pi
r(k_BT)^2
\over \ga_0 w^2}$ pushing it away from the daughter \rLiLiSa. The line
tension however creates a two-dimensional ``disjoining pressure,''
providing another
contribution to the effective free energy of $2\pi\ga_0 (r+w)$.
Minimizing the total free energy gives an average gap width of
$w=60\,$nm. A gap this wide certainly allows free and rapid diffusion
of small solutes like sugar.
Prior to the halfway point the pressure also helps close the
gap; after the halfway point nothing keeps the gap closed so that
expulsion  finishes rapidly.

While $w$ turns out to be larger than twice our proposed micelle
radius $2r_m\approx 10\,$nm, so that ultimately the micelles will
simply escape, nevertheless initially there will be an
osmotic pressure. Following the analysis of Finkelstein \rFink, let us
first neglect the line tension altogether and consider the free
flow at a leaky orifice with a concentration jump $\Delta
c$. We take the orifice to be a slit of length $L=2\pi r$ and width
$w\ll L$. There will be no pressure gradient down the center of the
slit, but the depletion zone of width $r_m$ will feel the van 't~Hoff
pressure jump $\Delta\pi\approx17\UP$ estimated earlier, leading to a
fluid velocity $v\approx {\Delta\pi\over\eta}r_m$ at the edge of the
depletion zone. Here $\eta=0.78\,$cP is the viscosity of water at
$31^\circ$C. Thus our approximate prediction for the flow rate is
$Q\approx Lwv\approx14$\muc/s, comparable to though somewhat less than
the observed
initial inward flow of about 100\muc/s (\fthree b). Our estimate of $Q$ could
be improved by accounting for hydrodynamic interactions, which
effectively increase the width of the depletion zone, and thus
increase the velocity and $Q$.

We  neglected the line energy $\ga_0$ in the
above estimate. The inward flow $Q$  just estimated will be partly
cancelled by an outward flow due to the interior pressure $p\approx
\ga_0/r^2 \approx1\UP$, reflecting the force needed to open the pore. The two
flows balance when $p=\Delta\pi{2r_m\over w}$ \rFink. Using our
estimate of $\Delta\pi$ we see that indeed the line tension is
unable to stop the inward flow.

\newsec{Conclusion}
Spontaneous
vesicle expulsion is a surprisingly complex behavior to emerge from
such a simple system, consisting of just water plus lipid. We have
seen how laser
tweezers act on lipid bilayers to create a simple
micron-scale machine, which pulls in material, repackages it, and
pumps water against a pressure gradient. Such machines are of
fundamental interest for the light they may shed on
processes in living systems. Expulsion also hints at the exciting
practical possibility of transforming
membrane structure when and where we wish to do so.
Further progress will require a better characterization of the objects
created by the laser, perhaps by light scattering from the
interior of the outer vesicle, and a more sophisticated theory, \eg\
treating transport through a fluctuating gap.

{\frenchspacing
\ifx\prlmode\testp\else\vskip1truein \leftline{\bf Acknowledgments}
\noindent\fi
We would like to thank
R. Bruinsma, N. Dan,
D. Fygenson, M. Goulian, R. Granek,  S. Gruner,
J. Israelachvili,
R. Kamien, M. Kraus,
N. Levit, A. Libchaber, R. Lipowsky, T. Lubensky,
R. Menes, S. Milner, C. Peskin,
S. Safran, U. Seifert, V. Suraiya,
and D. Zhelev
for their help. }
This work was supported in part by the Minerva Center
for Nonlinear Science, the Minerva Foundation (M\"unchen, Germany),
the US/Israeli Binational Foundation grant 94--00190, and NSF grant
DMR95--07366. EM was
supported in part by a Madeleine Haas Russell
Chair; JDM was supported in part by an FCAR Graduate Fellowship from
the government of Quebec.

\footatend\bigskip\immediate\closeout\rfile\writestoppt
  \baselineskip=22pt\centerline{{\bf References}}\bigskip{\frenchspacing%
  \parindent=20pt\escapechar=` \input refs.tmp\vfill\eject}\nonfrenchspacing
 \vfill\eject\immediate\closeout\ffile{\parindent40pt
 \baselineskip22pt\centerline{{\bf Figure Captions}}\nobreak\medskip
 \escapechar=` \input figs.tmp \vfill\eject
}

\ifx\figflag\figI\else\vfill\eject\immediate\closeout\ffile
\centerline{{\bf Figure Captions}}\bigskip\frenchspacing%
\input figs.tmp\vfill\eject\nonfrenchspacing\fi
\bye